\documentclass[prl,twocolumn,amsfonts,amssymb,amsmath,floatfix,tightenlines,superscriptaddress]{revtex4-1}

\usepackage{amsfonts,amssymb}
\usepackage{amsmath,times}
\usepackage{graphicx}
\usepackage{hyperref}    
\usepackage{color}
\usepackage{cleveref}
\usepackage[lofdepth,lotdepth,caption=false]{subfig}
\usepackage[normalem]{ulem}

\usepackage{booktabs}

\begin{document}

\title{
A Different Perspective on Superconductivity in Crystalline Graphene: Exploiting Energetics }

\author{Ke Wang}
\affiliation{Department of Physics and James Franck Institute, University of Chicago, Chicago, Illinois 60637, USA}
\author{Shicong Song}
\affiliation{Department of Physics, Florida Atlantic University, 777 Glades Road, Boca Raton, FL 33431-0991, USA}
\author{K. Levin}
\email{levin@jfi.uchicago.edu}
\affiliation{Department of Physics and James Franck Institute, University of Chicago, Chicago, Illinois 60637, USA}

\begin{abstract}
A central mystery of crystalline graphene is why superconductivity
is so widespread yet often confined to strange slivers near boundaries
between distinct isospin-ordered metals. In this paper, we apply a
``two-parent'' energetic framework which we show can explain this
unusual form of superconductivity \textit{without} specifying the
details of the necessarily present pairing attraction. First-order
transitions are crucial here: when two normal isospin-ordered states
are degenerate in free energy, even a small net superconducting
energy gain may stabilize an equilibrium superconductor. We
demonstrate how this is possible even though the small energy gain
from pairing is reduced by the expense of reconstructing the normal metal,
which is needed to achieve superconducting compatibility.
The first order degeneracy also gives superconductivity a choice between
two normal state parents, favoring the state with the largest net free
energy gain.
\end{abstract}

\date{\today}

\maketitle

\section{Introduction~}

Superconductivity in crystalline graphene is described as ubiquitous,
``a common feature''~\cite{Holleis2025}. It is found 
~\cite{Zhou2021,Zhou2022,Holleis2025,Han2025,Patterson2025,Yang2025,Seo2026,Auerbach2025,Kalantre2026,Dutta2026,DelaBarrera2022,Pantaleon2023}
in rich and complex phase diagrams competing with a variety of different
isospin ordered normal states~\cite{Zhou2021}
and commonly appearing in narrow regions near transitions
between two isospin orders.

This paper addresses these recurring features in the context of the
phase diagram represented schematically in Fig.~1. An essential first
step is to recognize that, in an itinerant system in which the same
electrons participate in both normal-state order and superconductivity,
the equilibrium phase must be established relative to multiple competing
electronic configurations. The pairing attraction is widely addressed
in the literature~
\cite{Chou2021,Ghazaryan2021,You2022,Cea2022,Chatterjee2022,Qin2023,
Dong2023,Dong2023b,Dong2023a,Li2023,Shavit2023,Dong2026}
but knowing it does not establish under what circumstances superconductivity wins the competition,

We address this complementary side of the problem using thermodynamics.
Such considerations determine where, once pairing is available,
superconductivity can stably compete with the surrounding
isospin-ordered metals. We develop a two-parent thermodynamic framework
and emphasize a central point: a first-order transition between two
isospin states qualitatively changes the energetic competition. At the
transition neither normal state has an energy advantage, so even a small
net superconducting gain can become decisive. The degeneracy also gives
superconductivity the flexibility to choose whichever normal state
provides the more favorable parent. There are antecedents 
~\cite{Anderson1973,Fradkin2015,Fernandes2010,Shavit2025,Wang2025}, 
emphasizing
thermodynamic competition in multi-ordered superconductors, but in
different physical settings.

\begin{figure}
    \centering
    \includegraphics[width=2.8in]{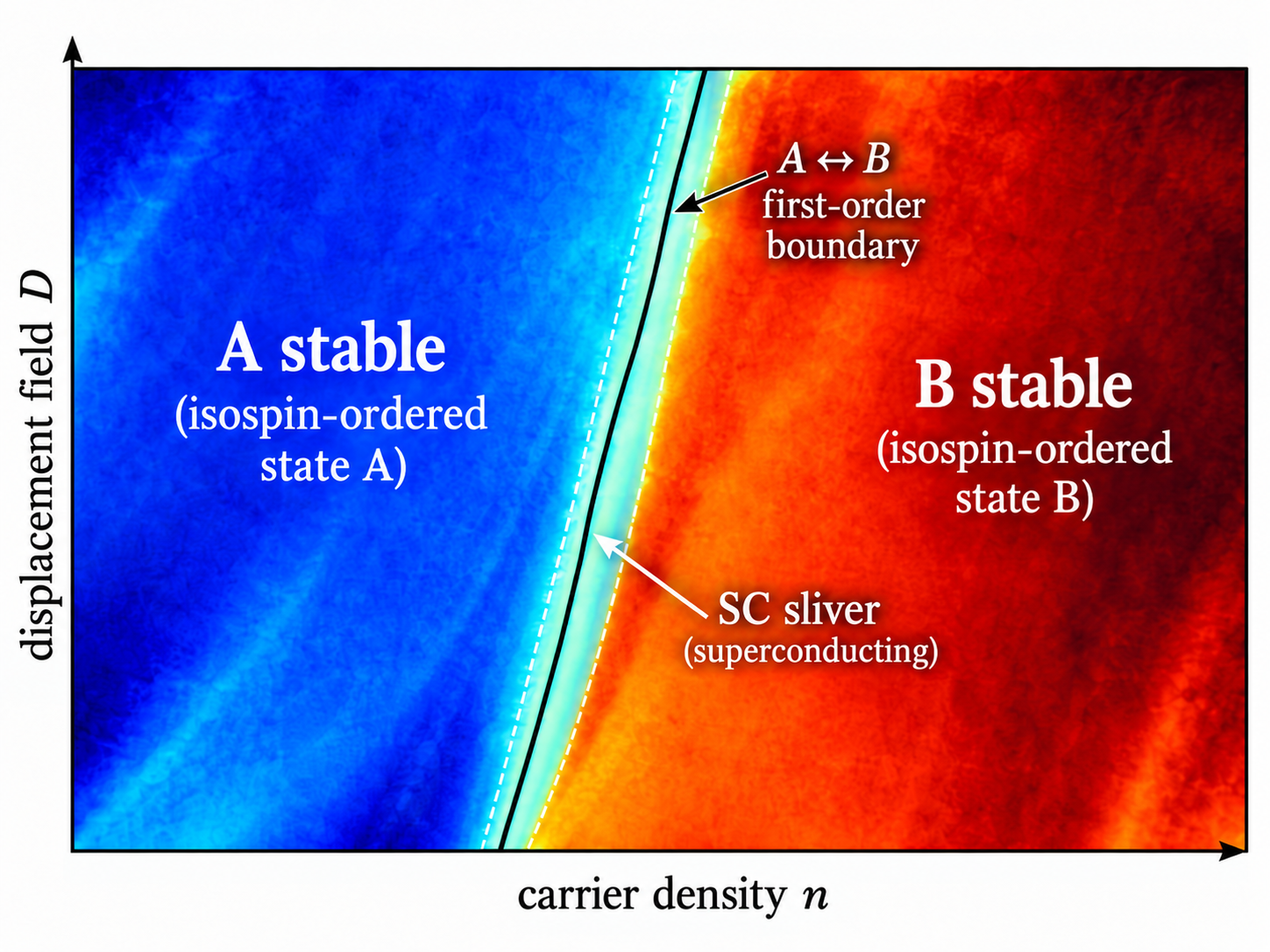}
\caption
{ Schematic (n)-(D) phase diagram near a first-order transition between two distinct isospin-ordered metallic states (A) and (B) as well as the superconducting sliver. The black curve
denotes the underlying normal-state $(A\leftrightarrow B)$ first-order boundary.
}
    \label{fig1}
\end{figure}

A consequence of this framework is the emergence of superconducting
slivers. We also find a natural asymmetry between the two sides of such
a sliver, which can be pronounced. 
Consequently, many slivers may be confined to one side or the other of the isospin boundary~\cite{Holleis2025}.
It is notable that we find superconductivity generally coexists
with isospin order. Conversely, the normal isospin ordered state has to reorganize
to accommodate pairing. This introduces a reconstruction cost which in turn
reduces the net superconducting gain. It is fortunate that the unusually rich landscape of
competing isospin states provides many opportunities making it possible for superconductivity
to, nevertheless, become competitive.

Finally, crystalline graphene also supports broader superconducting
regions~\cite{Holleis2025} rather than only narrow slivers. In these dome-like regimes, a
nearby first-order isospin transition is not generally required,
although reconstruction of the itinerant parent can still play an
important role.

\section{Two-parent energetic framework~}

 As noted above, an equilibrium treatment of an electronic superconductor must include the competing normal-state configurations as well as the superconducting state(s). We stress that the situation in rhombohedral
graphene is quite unusual as here two normal states can act as candidate parents of a superconducting state.
This, in turn, we argue, reflects the rarity of superconductivity experimentally appearing in the form of slivers. 
Here in our thermodynamics, we consider two distinct locally stable normal states, \(A\) and \(B\), whose free energies are \(F_A^N(x)\) and \(F_B^N(x)\), where \(x\) denotes a generic tuning parameter such as carrier density or displacement field. At a first-order transition, these are presumed equal at \(x=x_0\).

To make this explicit, let $M$ collectively denote the normal-state
electronic variables.  A minimal Landau free energy may be written as
\begin{equation}
{\cal F}(M,\Delta)
=
F_N(M)
+
a_{\rm eff}(M)|\Delta|^2
+
\frac{b}{2}|\Delta|^4,
\qquad b>0,
\label{eq:LGgeneral}
\end{equation}
where the dependence of $a_{\rm eff}$ on $M$ represents the
compatibility, or incompatibility, between the normal-state
configuration and superconductivity.
Here \(\Delta\) denotes the superconducting order parameter in whichever pairing channel is realized; its internal spin, valley, and orbital structure is left implicit, with the corresponding compatibility with the normal-state configuration absorbed into \(a_{\rm eff}(M)\), as well as other relevant variables.

The normal-state parents correspond to distinct minima, $M_A$ and $M_B$
with normal state energies denoted by
$F_{A,B}^N$.
For a sufficiently strongly ordered particle-hole parent $A$, it is possible that
superconductivity cannot be supported
$a_{\rm eff}(M_A)>0$, but it may
become favorable if the normal
configuration is allowed to reorganize.  Let $M_S^A$ denote the
configuration of the optimized superconducting state associated with
the $A$ configuration so that
$a_{\rm eff}(M_S^A)<0$.
The pairing energy gain at this reconstructed configuration is
$E_{\rm pair,A}
=
\frac{
a_{\rm eff}^2(M_S^A)
}{
2b
}
>0$.

Then one can define an important quantity which we call the reconstruction cost
\begin{equation}
\Delta F_{\rm rec,A}
\equiv
F_N(M_S^A)-F_N(M_A)
\geq 0.
\label{eq:FrecA}
\end{equation}

The free energy of the fully relaxed \textit{superconducting} state is 
\begin{equation}
F^{S}_{A}
=
F_A^N
+
\Delta F_{\rm rec,A}
-
E_{\rm pair,A} \equiv F_A^N -
E_{\rm SC,A}^{\rm net}
.
\label{eq:FSA}
\end{equation}
It is worth emphasizing that the reconstructed configuration $M_S^A$ of the normal particle-hole
states arises
from the fact that the same low-energy electrons participate in both the particle--hole order and Cooper pairing.

If we focus on the first-order boundary on which $A$ is
only metastable so that
$F_A^N>F_B^N$
then the reconstructed superconducting state becomes the equilibrium state
when
\begin{equation}
E_{\rm SC,A}^{\rm net}
>
F_A^N-F_B^N.
\label{eq:4}
\end{equation}

This is the central two-parent energetic criterion.  
There is a further consequence of the two-parent structure which is
important to emphasize.  At a first-order boundary,
superconductivity
may reconstruct from either parent configuration
%

At the first-order boundary the \textit{optimal} superconducting state is
determined from Eq.~(\ref{eq:4}) with
\begin{equation}
E_{\rm SC,opt}^{\rm net}
=
\max_{i=A,B}
\left[
E_{{\rm pair},i}-\Delta F_{{\rm rec},i}
\right].
\label{eq:5}
\end{equation}
At the
crossing, $F_A^N=F_B^N$, thus implying either normal configuration can
serve as the starting point for superconducting reconstruction without
having to assume a normal state free energy penalty.  
Note this first order boundary is not there to guarantee stronger attraction but rather it
serves to remove the 
normal state energy competition.

\begin{figure*}
    \centering
    \includegraphics[width=5.9in]{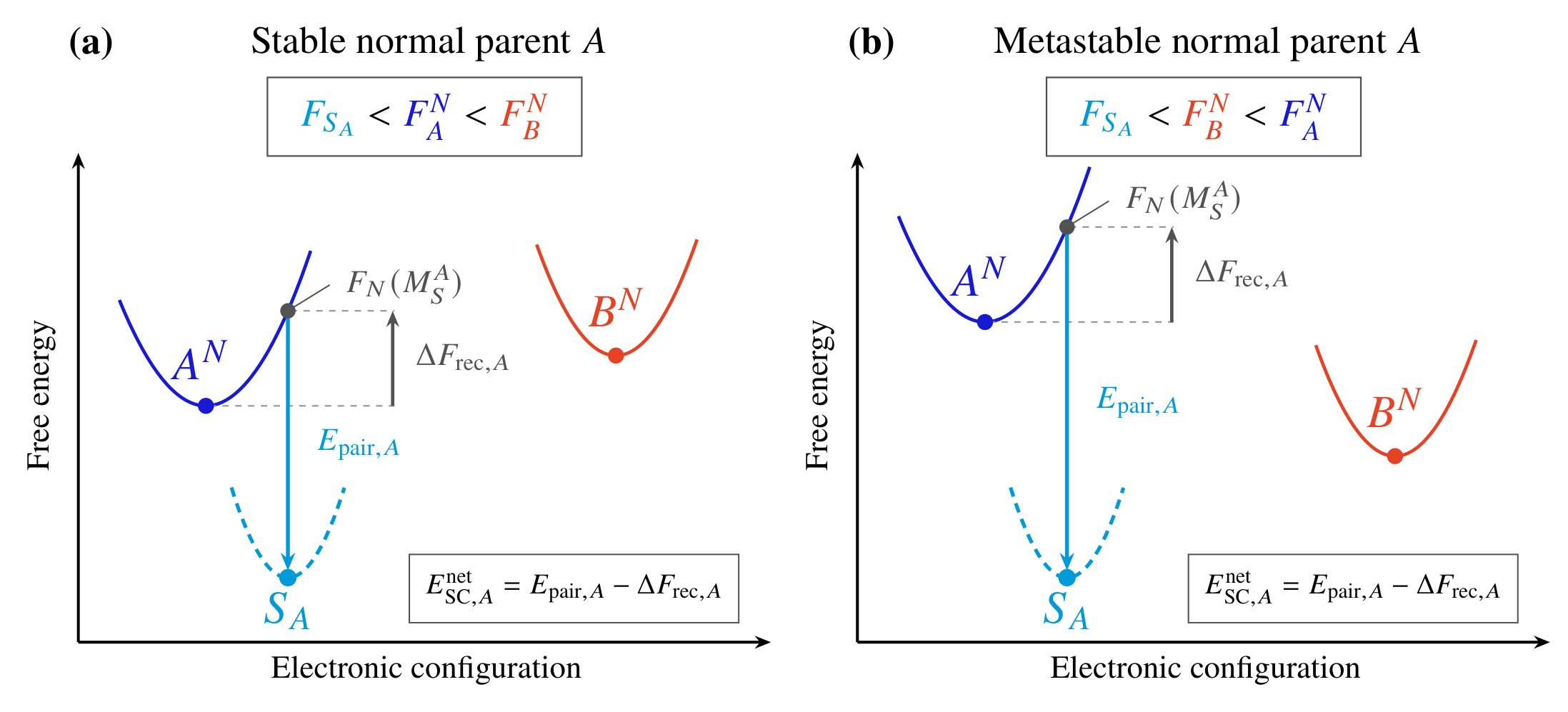}
\caption{Schematic free-energy landscapes for a superconducting
descendant of state $A$, called $\mathcal{S}_{A}$.  The horizontal axis is a collective
electronic-configuration coordinate, representing changes in the
spin, valley, isospin, or Fermi-surface reconstruction; it is not
an externally varied phase-diagram parameter.
The gray point is the normal-state energy of the configuration adopted by \(S_A\); its height above \(A^N\) is the reconstruction cost \(\Delta F_{\mathrm{rec},A}\).
At fixed external
parameters and low temperature, (a) $A$ is the stable normal
parent, whereas in (b) $A$ is metastable to $B^N$ but its
superconducting descendant $\mathcal{S}_{A}$ becomes the equilibrium
state.  In both panels, the cyan vertical arrow indicates the
pairing energy gain associated
with the $A$-derived superconducting state.}
   \label{fig1}
\end{figure*}


\section{Superconducting Slivers~}

As seen in Figure 2, the two relevant states can be categorized by stable-parent
and metastable parent.
We
want to first understand how slivers of superconductivity can descend from meta-stable state $A$.
This is addressed in Figure 2(b).

Let \(x\) denote a tuning parameter in the phase diagram, for example representing
density $n$ or displacement field $D$.  Define 
$\Delta F_A(x)
\equiv
F_A^N(x)-F_B^N(x)>0$.
The stability condition becomes
$0<
\Delta F_A(x)
<
E_{\rm SC,A}^{\rm net}(x)$.
Since $\Delta F_A(x_0)=0$ and 
$\Delta F_A(x)$
ordinarily grows upon moving away from
the first-order boundary, a small
$E_{\rm SC,A}^{\rm net}$ naturally confines the superconducting state
to a narrow region in parameter space.

Close to $x_0$,
\begin{equation}
\Delta F_A(x)
\simeq
\left.
\frac{\partial \Delta F_A}{\partial x}
\right|_{x_0}
(x-x_0),
\end{equation}
which gives the metastable-side estimate of the sliver size

\begin{equation}
\delta x_{\rm SC,A}
\sim
\frac{
E_{\rm SC,A}^{\rm net}(x_0)
}{
\left|
\partial_x\Delta F_A
\right|_{x_0}
}
\label{eq:5a}
\end{equation}
assuming that \(E_{\rm SC,A}^{\rm net}\) varies slowly across the narrow interval and that \(\partial_x\Delta F_A|_{x_0}\neq0\).

These estimates apply specifically to the extension of superconductivity onto
the metastable-parent side of the normal-state boundary.  When the free-energy
mismatch grows rapidly on that side, there may be little penetration of the
sliver across the isospin boundary.
This effectively leaves a one-sided superconducting
sliver confined predominantly to the stable-parent side.  This appears consistent
with the frequent observation of one-sided slivers~\cite{Guo2025}.

We now consider Figure 2(a) which is the side of the normal-state first-order boundary on
which $A$ is already the stable normal state.
This side is qualitatively different from the metastable-parent side.
Here the relevant competition is internal to the $A$ state.

Let the fully relaxed superconducting state associated with the
$A$ configuration have free energy
$F_{S_A}(x)
=
F_A^N(x)
+
\Delta F_{{\rm rec},A}(x)
-
E_{{\rm pair},A}(x)$,
where $\Delta F_{{\rm rec},A}$ is the normal-state cost of
reorganizing the $A$ configuration into one compatible with pairing,
and $E_{{\rm pair},A}$ is the superconducting energy gain obtained
after this reorganization.

On the stable-parent side, superconductivity can appear only when
the condensation energy is greater than the reorganization cost:
\begin{equation}
E_{{\rm pair},A}(x)
>
\Delta F_{{\rm rec},A}(x).
\label{eq:stablecriterion2}
\end{equation}
If the superconducting phase terminates within the stable-\(A\) region, its edge \(x_c^{(A)}\) is determined 
when the positive quantity $$ E_{\rm SC,A}^{\rm net}(x_c^{(A)}) \rightarrow 0. $$

Physically, as one moves deeper away from the transition and into the stable $A$ phase, the
normal-state order is expected to become increasingly well established and more
costly to reorganize.  At the same time, the pairing gain may weaken.
Either effect can drive $E_{{\rm SC},A}^{\rm net}$ to zero and thereby
terminate the superconducting state at $x_c$.

These issues are not unrelated to those discussed in Ref.~\cite{Wang2025} where 
for simpler magnetic superconductors it is
argued superconductivity emerges most strongly in the vicinity of weakening magnetism.
This latter can be interpreted as associated with minimal reconstruction costs. 
It is useful to make these comments more quantitative using Hartree Fock theory as the basis for 
typical free energy
plots on the metastable side.  We show the contrast between weakly and strong first order behavior in Appendix I.
In the weakly first order case there is more penetration across the isospin boundary whereas in the strongly
first order case the slivers are expected to be more one sided, as might be expected from
Eq.~(\ref{eq:5a}).

\section{Beyond Slivers to Domes: The One Itinerant Parent Case}

The superconducting phase diagrams of crystalline graphene are not
exclusively composed of narrow slivers.  An especially important
example is the higher-$T_c$ state SC2 in spin--orbit-coupled
Bernal bilayer graphene.  Holleis {\it et al.}~\cite{Holleis2025} find that SC2 develops
deep inside the nematic phase and, over most of its extent,
is not correlated with an adjoining isospin transition.  
There is an interesting claim made by these authors as they deduce from
their experiments that
the often observed proximity of
superconductivity to isospin phase boundaries in crystalline graphene
may actually be `` coincidental.''
While we would contest this point
of view, it is worthwhile to present it here,
Indeed, SC2 in Bernal graphene subject to spin-orbit coupling
provides a particularly clear realization of the one-parent
energetic case of the present paper.
In this way, the SC2 result demonstrates that proximity to a normal-state isospin boundary
is not a necessary condition for superconductivity.

Indeed, for a \textit{single} itinerant
parent the superconducting state is stable when
\begin{equation}
        E_{\rm pair}>\Delta F_{\rm rec},
\end{equation}
where $\Delta F_{\rm rec}$ is the reconstruction cost.
In contrast to the two-parent case,
there is no second isospin state invoved and
thus no thermodynamic reason for the
superconducting state to be confined to a narrow interval surrounding
a phase boundary.  A broad dome can occur wherever
$E_{\rm pair}(x)-\Delta F_{\rm rec}(x)>0$.

This distinction also clarifies the relation to conventional BCS
theory.  In the usual fixed-background treatment,
$\Delta F_{\rm rec}=0$, so that a Cooper instability which lowers the
free energy is sufficient.  In an itinerant ordered metal, by
contrast, the same low-energy electrons participate in the
particle--hole order as in Cooper pairing.
Pairing and thermodynamic
phase selection are then not independent issues. And this applies whether or not there are
one or two parent states.

\section{Experimental Evidence for First Order and First-Order-Like Transitions~}

We have stressed the importance of first order transitions as these are
a mechanism that enables the condensation energy to be effective in the energy balance.
Experimental normal-state cascade literature in Bernal and rhombohedral graphene
shows that related abrupt changes among spin-, valley-, layer-, and
Fermi-surface-polarized metals are a recurring property of these systems
~\cite{DelaBarrera2022,Seiler2022}.  Thus, superconductivity develops
within a phase space already populated by several closely competing
metallic phases, often separated by first order transitions.

In Appendix II we discuss more precisely how and where these first order effects
are observed and
note that there are theoretical studies~\cite{Raines2024}
which addressed in a more analytic framework
the origin of many of these strong first order transitions~\cite{Zhou2021}.
While the Table emphasizes the support for first order transitions, the clearest example of an \textit{exception} is the higher-\(T_c\) superconducting state SC2 in
the presence of Ising spin orbit coupling found in Bernal bilayer graphene \cite{Holleis2025}.  
This was discussed earlier.
The magnetic-field response of SC3 in rhombohedral
pentalayer graphene may provide another exception not cleanly associated
with a superconducting sliver.

\section{Effects of External Perturbations}

An especially striking feature of crystalline graphene is the
multiplicity of superconducting phases which appear or become enlarged
under magnetic field and proximity-induced spin--orbit coupling~\cite{Holleis2025,Shavit2023}.  This
raises a broader question: do these perturbations help only by
strengthening the microscopic pairing interaction, or can they also
make superconductivity more readily accommodated within an ordered
metal?

The narrow superconducting slivers suggest a delicate thermodynamic
balance between the paired state and neighboring ordered metals.  A
magnetic field or spin--orbit perturbation which is modest on the scale
of the underlying isospin ordering can nevertheless modify this balance
if it shifts competing states differently.  Such perturbations may
therefore move, expose, or enlarge superconducting regions.  The effect
is particularly favorable when the perturbation selects an electronic
configuration which is more compatible with pairing.

A magnetic field provides a particularly transparent example.  Near a
first-order boundary, the competing normal states are already close in
free energy, so that a relatively small magnetic energy can modify
their competition without overcoming the full scale of the isospin
ordering energy.  When the field preferentially stabilizes a parent
which better accommodates the superconducting state, its
superconducting descendant may become competitive where it was
previously preempted.

Ising spin--orbit coupling (SOC) provides a related route through its effect
on the spin--valley structure of the ordered metal.  SOC can change the
competition among polarized, spin-canted, intervalley-coherent, and
nematic states and thereby favor combinations of parent order and
superconductivity which can coexist at relatively low energetic cost.
Near a first-order boundary, the small inter-parent energy penalty makes
such changes especially consequential.

The same perturbation can also act within a single itinerant parent by
stabilizing an electronic configuration which is more compatible with
pairing and thereby reducing the reconstruction cost.  In this case no
nearby branch crossing is required.  Superconductivity can then remain
stable over an extended region satisfying
\begin{equation}
E_{\rm pair}-\Delta F_{\rm rec}>0,
\end{equation}
giving rise to a dome rather than a narrow sliver.  The SC2 state in
spin--orbit-coupled Bernal bilayer graphene provides a suggestive
example of this one-parent limit.

A useful direction for future experiments which would serve to support
or falsify the importance of first order physics stressed here, is to characterize
the normal state systematically under the same magnetic-field and SOC
conditions for which superconducting regions appear.  Measurements of
compressibility or chemical potential, quantum oscillations, magnetic
response, and, where possible, spin--valley structure could determine
whether superconductivity is accompanied by motion of a first-order
boundary, a change of normal-state parent, or an internal reconstruction
within an apparently unchanged ordered phase.

\section*{Conclusions}

New frontiers in electronic superconductivity should not be viewed as tied to crystalline graphene. 
The heavy-fermion system UGe$_2$~\cite{Saxena2000} and twisted bilayer WSe$_2$~\cite{Guo2026,Xia2026} provide additional examples of superconducting phase diagrams that may possibly be understood within the present energetic perspective.
These alternative superconductors, which will be explored elsewhere,
both provide settings in which superconductivity develops near competing magnetic or reconstructed electronic states
although the detailed phase-diagram geometries differ.

More generally, this paper lays the foundation for a better understanding of electronic superconductivity. 
While most would argue that superconductivity and the pairing medium (e.g., magnetic order)
compete in a rather straightforward way we are reminded here of something different. These two are all one system and so the competition is unavoidable in a thermodynamic sense. Even if we acquire a fully "successful" understanding of non-phonon glue-
thermodynamics is essential. What is
unusual about crystalline graphene is its unique phase diagram. It is hard to come up with
another case where there is a random scattering of superconducting slivers-- even in
electronic superconducting systems like heavy fermions or pnictides. 
Crystalline graphene has provided a whole new laboratory for learning about electronic superconductivity.

\vskip5mm
\textit{Acknowledgement.}
We warmly thank Zhiqiang Wang for providing some of the inspiration for this work
through our previous collaborations.
We thank Long Ju, Haoxin Zhou, A. Sharpe, and Ludwig Holleis for reading the manuscript and their useful comments.
We also acknowledge the University of Chicago's Research Computing Center.

\bibliography{References2}

\clearpage

\twocolumngrid

\begin{widetext}

\begin{figure}
    \centering
    \includegraphics[width=6in]
{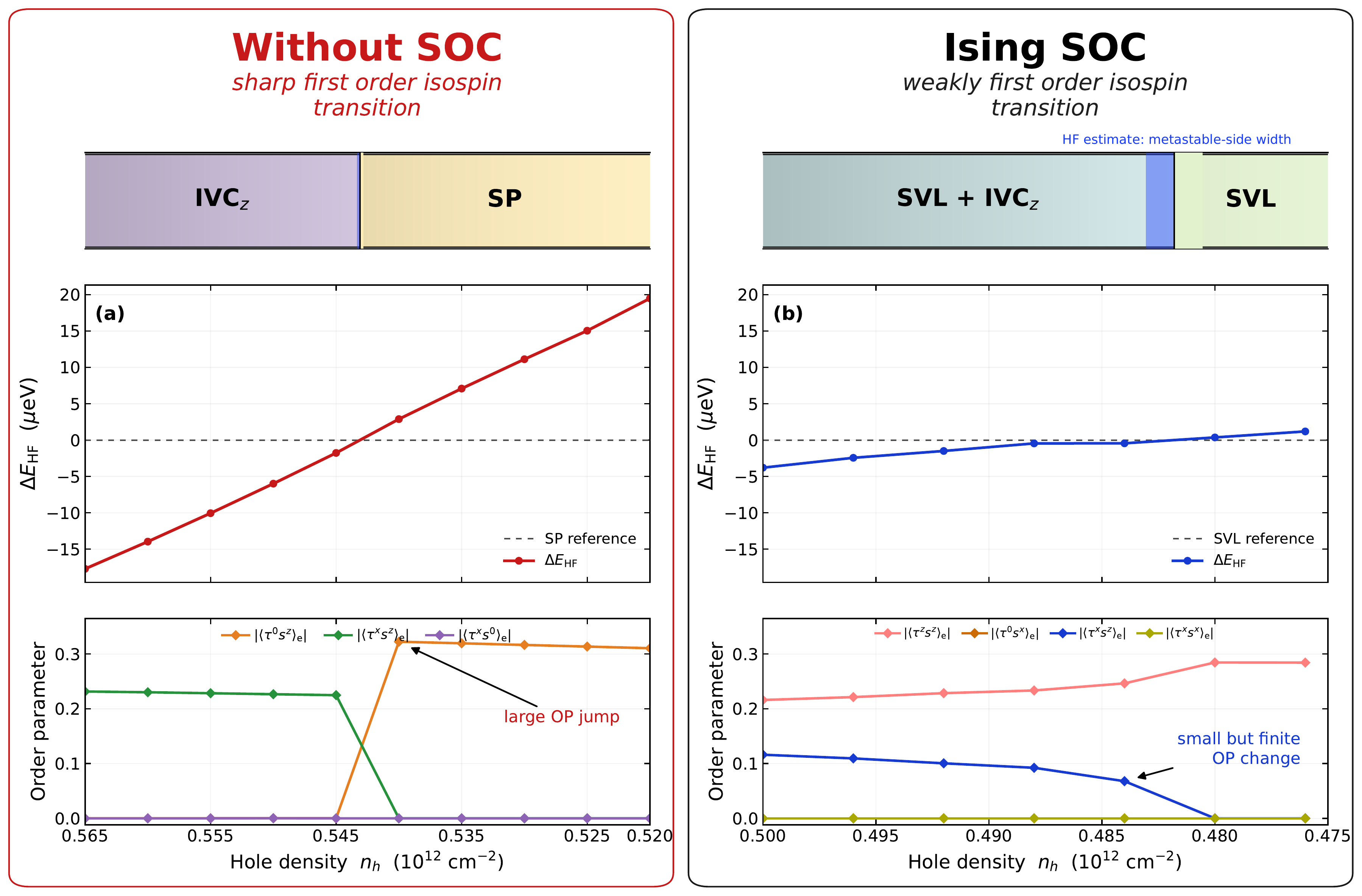}
\caption{
Normal-state Hartree--Fock comparison relevant to the metastable-side
extension of an $A$-derived superconducting state.
Left: in the absence of SOC, the transition between the
$\mathrm{IVC}_z$ and spin-polarized (SP) states is accompanied by a
relatively steep crossing of the Hartree--Fock energies and a large
discontinuous change in the isospin order parameters, characteristic of
a sharp first-order reconstruction.
Right: with Ising SOC, the competing
$\mathrm{SVL}+\mathrm{IVC}_z$ and SVL states remain separated by a
first-order transition, but the normal-state energy splitting grows much
more slowly and the order-parameter discontinuity is correspondingly
weak.
The blue band is drawn only on the metastable-parent side and
schematically represents the extension
$\delta x_{\mathrm{SC},A}^{\mathrm{meta}}$ estimated from Eq.~(7).
The extent on the stable-parent side, governed by
$E_{\mathrm{pair},A}=\Delta F_{\mathrm{rec},A}$, is not calculated here.
The two panels compare representative transitions and are not intended
to show continuous evolution of one boundary into the other.
Here \(A\) denotes the right-hand SP state in the left panel and the
right-hand SVL state in the right panel; the blue region represents
the continuation of this parent into the region where it is metastable.
}
    \label{fig4}
\end{figure}

\section*{APPENDIX I: HARTREE--FOCK NUMERICS:
FREE-ENERGY EVOLUTION ON THE METASTABLE SIDE}

The principal purpose of this appendix is to illustrate how the
normal-state free-energy landscape controls the extension of a
superconducting state onto the metastable-parent side of a first-order
boundary.  This provides a possible explanation for why experimentally
observed superconducting slivers can appear strongly one-sided.

Consider an $A$-derived superconducting state entering the region in
which $B$ is the stable normal state.  The relevant stability condition is
\begin{equation}
E^{\rm net}_{\rm SC,A}(x)
>
F_A^N(x)-F_B^N(x),
\end{equation}
and, close to the first-order crossing $x_0$,

\begin{equation}
\delta x_{\mathrm{SC},A}^{\mathrm{meta}}
\sim
\frac{E_{\mathrm{SC},A}^{\mathrm{net}}(x_0)}
{\left|\partial_x\!\left(F_A^N-F_B^N\right)\right|_{x_0}} .
\label{eq:meta_width_appendix}
\end{equation}

Thus the rate at which the two normal-state free energies separate away
from the crossing has a direct physical consequence.  If
$F_A^N-F_B^N$ rises steeply on entering the $B$-stable region, the
superconducting descendant of $A$ can penetrate only a very short
distance into the metastable territory.  The resulting superconducting
region may then appear experimentally to lie almost entirely on the
stable-$A$ side of the normal-state boundary.  By contrast, when the
normal-state free-energy splitting grows only slowly, the same net
superconducting gain can sustain superconductivity over a substantially
larger interval on the metastable side, and the sliver need no longer
appear strongly one-sided.

Figure~3 in the text compares two representative constant-\(g\) transitions in Bernal bilayer graphene: \({\rm IVC}_z\leftrightarrow{\rm SP}\) without SOC and \({\rm SVL+IVC}_z\leftrightarrow{\rm SVL}\) with Ising SOC.
The comparison is not intended as a continuous tracking of the same phase boundary as SOC is turned on.
Rather, both transitions preserve \(g=2\), thereby avoiding the more obvious flavor-depletion transitions in which the number of occupied spin-valley flavors changes. They therefore provide a useful comparison of how strongly two same-\(g\) isospin states can compete in the two landscapes.

Illustrated are these two possibilities using representative
normal-state Hartree--Fock calculations.  In the absence of SOC, the
transition between the competing states is accompanied by a relatively
steep crossing of the Hartree--Fock energies.  For a fixed
$E^{\rm net}_{\rm SC,A}$ this would imply only a narrow penetration of
an $A$-derived superconducting state into the metastable-parent region.
With Ising SOC, the representative weakly first-order transition shows
a much slower growth of the normal-state free-energy splitting.  The
corresponding metastable-side extension can therefore be substantially
larger.

It is important that this comparison constrains only the
metastable-parent side of the superconducting region.  The extent on
the stable-$A$ side is governed by a different balance,
\begin{equation}
E_{\rm pair,A}(x)
=
\Delta F_{\rm rec,A}(x),
\end{equation}
and cannot be obtained from the normal-state Hartree--Fock energy
splitting alone.  The blue regions in Fig.~3 consequently represent
only the schematic metastable-side extensions inferred from
Eq.~(\ref{eq:meta_width_appendix}), rather than the full widths of the
superconducting slivers.

An additional implication of this comparison concerns spin--orbit
coupling.  Ising SOC can introduce weakly first-order boundaries for
which the competing free-energy branches separate relatively slowly.
Such boundaries can make the metastable-side portion of a
superconducting sliver more readily visible, without requiring any
assumed enhancement of the microscopic pairing attraction.  We view
this as a secondary consequence of Fig.~3; its primary role is to show
how the normal-state free-energy evolution can determine whether a
two-parent superconducting region appears effectively one-sided.

\onecolumngrid

\section{Appendix II: First-Order Experimental Evidence}

%

\begin{table*}[!htbp]
\centering

{\normalsize\bfseries
Established first-order transitions near superconductivity
\par}
\vspace{6pt}
\small
\setlength{\tabcolsep}{4pt}
\renewcommand{\arraystretch}{1.15}

\begin{tabular}{@{}
p{0.18\textwidth}
p{0.23\textwidth}
p{0.30\textwidth}
p{\dimexpr0.29\textwidth-6\tabcolsep\relax}
@{}}
\toprule

\raggedright System or regime
&
\raggedright Normal-state transition
&
\raggedright Evidence and source
&
\raggedright Relation to superconductivity
\tabularnewline

\midrule

\raggedright Bare rhombohedral trilayer graphene
&
\raggedright Isospin-unpolarized metal $\leftrightarrow$
partially isospin-polarized metal
&
\raggedright Simultaneous transport and inverse-compressibility
measurements identify a first-order boundary through a
negative-compressibility peak~\cite{Holleis2025}, Fig.~4(a).
&
\raggedright Superconductivity lies immediately on the
isospin-unpolarized side of the measured boundary;
discovery context: Ref.~\cite{Zhou2021}.
\tabularnewline

\addlinespace[6pt]

\raggedright Bernal bilayer graphene at high $D$
and finite $B_\parallel$
&
\raggedright Isospin-reconstructing boundary adjoining
the superconducting ordered metal
&
\raggedright Simultaneous transport and inverse-compressibility
measurements identify the adjacent transition as first
order~\cite{Holleis2025}, Fig.~4(b).
&
\raggedright Field-induced superconductivity lies immediately
on the isospin-ordered side;
discovery context: Ref.~\cite{Zhou2022}.
\tabularnewline

\bottomrule
\end{tabular}

\caption{
Established first-order normal-state transitions adjacent to
superconductivity. The assignments follow the boundary-specific
measurements and analysis of Ref.~\cite{Holleis2025}, rather than
negative compressibility alone. The later trilayer scan is not
assigned one-to-one to every SC1/SC2 label of
Ref.~\cite{Zhou2021}.
}
\label{tab:established}

\end{table*}

\begin{table*}[!htbp]
\centering
{\normalsize\bfseries
First-order-like reconstructions near superconductivity
\par}
{\small\itshape
Thermodynamic order of the transition remains unresolved
\par}
\vspace{6pt}
\small
\setlength{\tabcolsep}{4pt}
\renewcommand{\arraystretch}{1.15}

\begin{tabular}{@{}
p{0.18\textwidth}
p{0.23\textwidth}
p{0.30\textwidth}
p{\dimexpr0.29\textwidth-6\tabcolsep\relax}
@{}}
\toprule

\raggedright System or SC region
&
\raggedright Observed normal-state change
&
\raggedright Evidence and source
&
\raggedright Relation and evidential limit
\tabularnewline

\midrule

\raggedright WSe$_2$-supported Bernal bilayer graphene, SC1
&
\raggedright SOC-split $\mathrm{Ising}_{2,6}$ metal
$\leftrightarrow$ nematic $N_{2,4}$ metal
&
\raggedright Quantum-oscillation frequencies change abruptly
at the low-$|n_e|$ edge of SC1~\cite{Holleis2025}, Fig.~2.
Extended Data Fig.~10(c) gives a noise-limited bound of about
$k_B(0.30\,\mathrm{K})$ on an unresolved chemical-potential jump.
&
\raggedright SC1 lies on the $\mathrm{Ising}_{2,6}$ side,
adjoining $N_{2,4}$. The bound is not a detected jump;
the transition order remains unresolved.
\tabularnewline

\addlinespace[6pt]

\raggedright Rhombohedral tetralayer graphene,
candidate chiral SC region
&
\raggedright Circular quarter-metal spectrum
$\rightarrow$ reconstructed ``multitone'' spectrum
&
\raggedright The single-tone quarter metal abruptly
disappears on reducing density. The multitone spectrum
persists throughout the SC region~\cite{Kalantre2026},
Figs.~3--4; discovery context: Ref.~\cite{Han2025}.
&
\raggedright The reconstruction occurs at or above the
SC onset density. The identity of the multitone state
and the thermodynamic order of the transition remain
unresolved.
\tabularnewline

\addlinespace[6pt]

\raggedright Rhombohedral pentalayer graphene,
chiral SC region
&
\raggedright Quarter-metal parent develops a domain-rich
texture; an additional hidden order is proposed
&
\raggedright Domain-wall proliferation above $T_c$ occurs
at the low-temperature SC onset density, accompanied by
rapidly reduced coercive fields and history-dependent
domains~\cite{Dutta2026}, Figs.~2--3;
discovery context: Ref.~\cite{Han2025}.
&
\raggedright These observations motivate a hidden
parent-state transition. Its microscopic identity and
thermodynamic order are unresolved; domain hysteresis
alone does not establish a first-order density-tuned
transition.
\tabularnewline

\bottomrule
\end{tabular}

\caption{
First-order-like reconstructions and supporting domain
observations near superconductivity. These are distinguished
from the established first-order boundaries in
Table~\ref{tab:established}. Abrupt fermiology and
magnetic-domain hysteresis do not by themselves determine
the thermodynamic order of a density- or
displacement-field-tuned transition.
}
\label{tab:unresolved}

\end{table*}


\end{widetext}

\end{document}